# Connected Vehicle Supported Adaptive Traffic Control for Near-congested Condition in a Mixed Traffic Stream

Sakib Mahmud Khan[1*], Mashrur Chowdhury[2]

[1]PhD Candidate, Glenn Department of Civil Engineering, Clemson University, Clemson, SC, 29634, USA; email: sakibk@g.clemson.edu

[2]PhD, Professor, Glenn Department of Civil Engineering, Clemson University, Clemson, SC, 29634, USA; email: mac@ clemson.edu

Abstract

Connected Vehicles (CVs) have the potential to significantly increase the safety, mobility, and environmental benefits of transportation applications. In this research, we have developed a real-time adaptive traffic signal control algorithm that utilizes only CV data to compute the signal timing parameters for an urban arterial in the near-congested condition. We have used a machine learning based short-term traffic forecasting model to predict the overall traffic counts in CV-based platoons. Using multi-objective optimization technique, we compute the green interval time for each intersection using CV-based platoons. Later, we dynamically adjust intersection offsets in real-time, so the vehicles in major street can experience improved operational conditions compared to loop-detector based actuated coordinated signal control. Using a 3-mile-long simulated corridor of US-29 in Greenville, SC, we have evaluated our CV-based adaptive signal control's performance. For the next time interval, using only 5% CV data, the Root Mean Square Error of the machine learning based-prediction is 10 vehicles. Our analysis reveals that the CV-based adaptive signal control improves operational conditions in the major street compared to the actuated coordinated scenario. Also, using only CV data, the operational performance improves even for a low CV penetration (5% CV), and the benefit increases with increasing CV penetration. We can provide operational benefits to both CVs and non-CVs with the limited data from 5% CVs, with 5.6% average speed increase, and 66.7% and 32.4% reduction in average maximum queue length and stopped delay, respectively, in major street coordinated direction compared to the actuated coordinated scenario in the same direction.

1. Introduction

Connected Vehicles (CVs) are wirelessly connected to mobile nodes that communicate with the surrounding connected vehicles, other connected road users (e.g., pedestrian, cyclists) and infrastructure (i.e., traffic signal, roadside unit) using any available communication option, and provide data based on vehicle movement, and interactions with the surrounding environment. Although many research-based or market-focused studies have predicted the rapid penetration of CVs into the regular vehicle fleet, challenges exist to develop an efficient transportation system considering both CV and regular vehicles (non-CVs). In this research, we have considered the mixed traffic scenario while developing an adaptive traffic control algorithm for the urban arterials. In urban arterials, we use the traffic signal control to control and manage traffic operations. The signal controller ensures the smooth traffic operations by reducing the delay and conflicts on urban arterials, and the controller collects data from different traffic data collection sensors or detectors. Widely used traditional traffic data collection sensors include inductive loop detectors, video camera, RADAR, and/or ultrasonic sensors. However, the emerging CVs have the potential to provide the data required by the traffic signal controllers, and, thus, it can help to reduce or even eliminate the need of traditional traffic sensors. In this research, we have used only CV based data to collect traffic flow information on the corridor, estimate existing queue length near the signalized intersections, and estimate the traffic signal timing parameters.

Adaptive traffic signal control is the most advanced strategy for signalized intersection operation and it promises to offer better operational, safety, environmental and economic performance compared to the fixed/pre-timed and actuated traffic signal control. Adaptive traffic signal control algorithms collect real-time data and they dynamically estimate traffic signal timing parameters satisfying different objectives (e.g., reduction of travel time, delay, and/or queue length, or increase in average speed), which ensures better performance compared to the actuated and pre-timed traffic control. Using data only from a limited number of CVs in a mixed traffic stream, estimating traffic signal timing parameters is inherently challenging as complexities can arise while estimating aggregated data (e.g., total vehicle count, average queue length, average speed) from microscopic data from CVs that are small proportion of the total traffic streams. An earlier study conducted by (Goodall, Smith, and Park 2014) could not achieve better operational performance with limited CV penetration (i.e., less than 25% CV) compared to coordinated actuated traffic signal control. Another challenging aspect of traffic signal design for an urban arterial is the establishment of proper signal coordination. Signalized intersections are closely spaced in urban areas; therefore, the traffic signals are often coordinated to make the vehicle progression smooth in the coordinated direction. Offset is one of the most important parameters for the signalized intersection coordination. Traffic engineers use the term 'offset' to refer to the time lapse between the start of green times between two successive coordinated, signalized intersections. Offset is a common parameter for actuated coordinated signals; however, most existing actuated coordinated signals have fixed offset values. During rush hour traffic, if these is a huge fluctuation in directional traffic flow, having fixed offset values may not provide the desired operational benefits due to the varying traffic demands. Another concern for the existing adaptive signal control algorithms is that the green signal interval does not update in real-time based on the traffic demand in the coordinated direction (Shelby et al. 2008). Earlier studies have estimated a fixed green signal interval for the intersections in the coordinated direction. However, during a complete cycle time (i.e., time required to complete all phases in an intersection), traffic volume can fluctuate in the coordinated direction; thus, having a volume-responsive green interval time can help to dynamically adjust green time for the coordinated direction in response to the existing demand.

In order to overcome the aforementioned limitations, we have developed a CV-based adaptive traffic signal control algorithm with three distinct capabilities. First, the algorithm can estimate traffic signal timing parameters based on the limited CV data in a mixed traffic environment. Second, the algorithm can conduct dynamic offset adjustments based on the existing congestion condition in signalized intersections. Third, the algorithm can do dynamic adjustment in the green time interval for the coordinated directions based on the traffic demand in corresponding directions. We have used a machine learning based short-term traffic forecasting model to predict the future traffic counts. Based on the predicted counts, we have identified vehicle platoons going through the intersections, and used a multi-objective optimization framework with Mixed Integer Linear Programming (MILP) models to estimate signal interval time for each intersection. We have extended the major street green intervals based on the available cycle time if there is no call from the minor street. At the last step, we have used another Mixed Integer Non-Linear Programming (MINLP) model to dynamically estimate the offset values for each signalized intersection. We have evaluated the CV-based adaptive signal control's performance using a simulation network of US 29, Greenville SC. Our motivation is to improve operational performance on the coordinated major streets with limited CV data. We have discussed the related study, the CV-based adaptive signal control algorithm, experimental design, analysis and discussion, and conclusion in the following sections.

2. Related study

In the following subsections, we have discussed the related studies on short-term traffic prediction, CV-based adaptive signal control, and traffic signal coordination.

*2.1. Short-term traffic prediction and platoon identification*

Common input variables for any vehicle-responsive traffic control systems (including the adaptive traffic signal control system) include vehicle arrival time, traffic count, queue length, time gap between vehicles, and detector occupancy (Gershenson 2004; He, Head, and Ding 2012; Ki et al. 2017; Xie et al. 2011; Yulianto 2018). We have used traffic counts as an input parameter for the adaptive traffic control system in this study. In addition, we have used a short-term time-series forecasting model to predict the traffic count in the future based on the existing CV data. In a review study by (Vlahogianni, Karlaftis, and Golias 2014), the authors identified challenges associated with the short-term traffic forecasting models. The traffic forecasting models for arterials are found to be more complicated than that of freeways, as traffic signals have a direct impact on arterial operations. Using data from a low penetration level of probe vehicles and identifying the proper data aggregation interval for forecasting are the two main challenges for traffic forecasting, which we have addressed in our study. Different state of the art approaches including the Auto Regressive Integrated Moving Average or ARIMA model, state-space models, and univariate and multivariate methods have been used by (Chatfield 2005) for time series prediction. An earlier study has established machine learning based models as viable options for short-term traffic forecasting (Vlahogianni, Karlaftis, and Golias 2014). Among the machine learning based models, the Recurrent Neural Network (RNN) model is one of the widely used models for sequence prediction. The layers of the neural networks have internal feedback connections, which are used to update the network weights until the model converges to reduce the difference between the actual data and predicted data. In their use of Long Short-Term Memory (LSTM) to predict traffic flow at a specific detector location, the authors in (Kang, Lv, and Chen 2018) found that the inclusion of speed and occupancy from upstream and downstream locations can improve prediction accuracy. In another study, (Tian and Pan 2015) found the LSTM model outperforms the Random Walk, Feed Forward Neural Network, Support Vector Machine and Stacked Auto-encoder to predict traffic flow for different prediction intervals.

*2.2. CV based adaptive traffic signal control*

Different studies have investigated the CV-enabled adaptive traffic control systems (Ban and Li 2018; Beak, Head, and Feng 2017; Feng et al. 2015; Goodall, Smith, and Park 2014; Guo, Li, and (Jeff) Ban 2019; Li and Ban 2018). In (Goodall, Smith, and Park 2014), the authors used simulation to predict traffic over a 15-second time horizon, and estimated signal timing for a decentralized adaptive signal that satisfied objective functions (both single-variable and multivariable objectives) for the predicted traffic. The single-variable objective function was the minimization of cumulative vehicle delay for the simulated corridor with four intersections in Virginia. With 50% or more penetration levels, delay reduction and speed improvements occurred. For the adaptive signal system, when the degree of saturation was lower than 0.9 and CV penetration was 100%, the operational performance improved or was not significantly altered compared to the coordinated actuated signal control. Using multi-variable objective function (i.e., with delay, negative acceleration, and number of stops), the operational performance did not improve compared to the single-variable objective scenario. (Feng et al. 2015) evaluated their algorithm in a simulation environment where the signal controller estimated position and speed of the non-CVs in three specific regions. Non-CVs were identified in the queuing, slow-down and free-flow regions using queue propagation speed, relative acceleration difference of CVs, and number of CVs with CV penetration rate, respectively. For left-turn lanes, a stop line detector was used to detect left-turning vehicles. Once information about both CVs and non-CVs were known, the controller predicted vehicle arrival times and used two-level optimization (with two separate objectives: to minimize total vehicle delay or minimize queue length) of the vehicles for a prediction horizon of 80 seconds. Simulation results showed that reductions in average delay for each phase occurred with 50% or more CVs for both objective functions in one intersection. (Beak, Head, and Feng 2017) extended the study of (Feng et al. 2015) by including coordination strategy (for a corridor with five signalized intersections) with the intersection-level control. The authors in (Beak, Head, and Feng 2017) developed a platoon flow model based on the platoon dispersion model and simulation generated data. The link performance function used the platoon flow model as an input. Using data from both CVs and the stop bar detectors, adaptive signal control reduced the average delay and average number of stops by 6% and 3%, respectively, in the coordinated direction with 25% CV penetration compared to the actuated coordinated control.

For the intersection-level traffic signal control, (Li and Ban 2018) used a dynamic programming model to calculate signal timing using CV speed and location data. In order to ensure fixed cycle length using dynamic programming, the authors used a penalty function and a branch-and-bound method. The authors developed a framework to reduce fuel consumption and travel time for an environment with 100% CV penetration. For different traffic demands (250, 650 and 800 vph), the dynamic programming-based signal control model outperformed the performance of the SYNCHRO based actuated traffic control, but did not outperform the MATLAB NOMAD-solver based signal control. However, the authors suggested to use the dynamic programming based signal control as it was more computationally feasible than the NOMAD-solver based control. In their review study, (Guo, Li, and (Jeff) Ban 2019) discussed two approaches used by researchers while developing signal control frameworks, which are: deterministic (i.e., shockwave and kinetic equation based approach) and stochastic approaches (i.e., uniform or Poisson distribution, machine learning) to derive traffic flow measures. Vehicle trajectory estimation is gaining more momentum to improve traffic signal performance. Based on the variations of prediction data, three different control methods were identified: (1) CAV-enabled actuated signal control, (2) platoon-based signal control, and (3) planning-based signal control (i.e., using data from each vehicle, predict the traffic state for a future time horizon, and optimize traffic signal time for that horizon).

For traffic signal timing estimations, creating platoons can help to reduce the computation complexities of a signal timing algorithm. Platoon, representing a group of vehicles, is a form of a disaggregated unit that the signal timing algorithm can consider while assessing vehicle progression. The authors in (Bashiri and Fleming 2017) discussed a platoon-based scheduling strategy for such scenario where all vehicles were Connected Automated Vehicles or CAVs, and platoons communicated with the intersection controller. When connectivity is incorporated with the automated/autonomous vehicles (i.e., vehicles having partial automation/full automation), the vehicles are referred to as Connected Automated Vehicles (CAVs). In an earlier study, once the controller knew the platoons arrival sequence , it implemented a greedy algorithm to reduce the waiting time of the arriving platoons (Bashiri and Fleming 2017). For a four-legged intersection, the controller with platoon based delay minimization objective achieved almost 50% less delay compared to the stop controlled intersection for vehicle flows ranging between 500 to 2000 veh/hour.

*2.3. Traffic signal control coordination*

(C. M. Day et al. 2017) considered CV trajectory data in a mixed traffic scenario to optimize the intersection offsets. With 1% CV penetration level, the authors constructed a vehicle arrival profile from the CV data using virtual detectors (i.e., hypothetical detectors to identify CVs), and compared the findings with the physical detector-based method (i.e., inductive loop detector to identify all vehicles). Using an earlier offset optimizer algorithm developed in (C. Day and Bullock 2012), the authors computed the offset for both virtual and physical detectors, and implemented the offsets separately for one week time period. Using CV-based virtual detectors, the authors achieved similar operational performance. Using sensitivity analysis, at a 90% confidence level, the authors found that a 2 week sampling period with the CV-based virtual detector method provided a similar vehicle arrival profile with the physical detector-based method. For offset optimization in urban areas, (Aoki, Niimi, and Kamijo 2013) considered arriving vehicles' profiles and traffic signal timing parameters (i.e., cycle length, split and offset values from adjacent intersections) to maximize the green time overlap. At first, the authors estimated the phase start and end time for the next cycle based on the phase end time of the current cycle, current phase split and reference cycle length. Later, they constructed the vehicle arrival profiles and using the profiles they maximized the green overlap time. Using different traffic demands, the green overlap maximization with vehicle arrival profiles was found to reduce the number of stops and delay for a grid network with 25 intersections, when compared with the Split Cycle Offset Optimisation Technique (SCOOT) and decentralized-SCOOT techniques.

(Zhang et al. 2016) developed a probe trajectory-based real-time offset estimation method based on vehicle state information (i.e., queue dissipation, queue formation, free flow). Using vehicle speed and location data, the free flow area (i.e., area where vehicle speed does not change more than 10% of the historical free flow speed), queue forming area (i.e., area starting from the point where vehicle speed is less than 3 mph) and queue dissipation area (i.e., area starting after the point where vehicle speed is less than 3 mph) are detected. The optimal offset is calculated as the difference between the time required by a free-flowing vehicle to traverse the corridor, and the time to dissipate the standing queue in the downstream intersection. For a simulated corridor with two intersections, the probe-vehicle based offset tuning method reduced the average delay under fluctuating traffic while compared with the model without offset tuning. But when there was no fluctuation in the traffic demand, the average delay did not improve with the real-time offset tuning method. (Jiao, Wang, and Sun 2014) developed a real-time coordinated signal control system for urban arterials by estimating turning flows and cycle length at each intersection, and finally calculating split and offset for the arterials. Using loop detectors, the authors collected entering and exiting vehicle counts, and, using those counts, estimated the intersection cycle time to minimize the average delay.

For a simulated corridor with three intersections, the real-time coordinated control outperformed the existing MAXBAND method (i.e., method to maximize the green wave band) by generating 16.7% less mean queue length, 6.4% less mean delay and 7.7% less number of stops. (Daganzo, Lehe, and Argote-Cabanero 2018) used only average traffic density to optimize the signal offsets for one-way streets while keeping the intersection green ratios the same. If the target corridor's density was greater than the optimum density, backward progression (i.e., adjusting offsets based on the backward moving observer) was used. If the density was lower than the optimum density, forward progression (i.e., adjusting offsets based on the forward moving observer) was used. The real-time offset optimization method generated better operational performance with large roadway sections, compared to short corridors. However, the authors created a hypothetical simulation environment (e.g., one-way street with uniformly spaced intersections), which was favorable for adaptive offset optimization. Further investigation in the complicated, real-world environment is needed to validate the framework for real-world implementation.

3. Research considerations for this study

In this study, we have developed a CV-based adaptive traffic control system for urban arterials. It is assumed that the urban arterial is instrumented with wireless communication devices (i.e., Roadside Units or RSUs) with computation capabilities required to estimate signal parameters in real-time. Urban arterials are characterized by multiple, closely spaced intersections, and multiple RSUs are required to establish proper coverage to receive real-time CV data. CVs will generate Basic Safety Messages (BSMs) which include time, location, speed, direction etc. BSMs are defined by the standards of Society of Automotive Engineers, specifically by SAE j2735 which defines the message structure, data frame and data element of BSMs (SAE 2009). We have used these CV generated BSMs to estimate the signal timing parameters. In addition, we have not considered the effect of driveway traffic on the mainline traffic in the case study.

4. CV-based Adaptive Signal Control Algorithm

The CV-based adaptive signal control algorithm follows three steps. First, we identify the CV-based platoons in both the major and minor directions to estimate the required green time intervals. Second, once we have an understanding about the approaching platoons, we compute the signal timing parameters for each intersection. Third, while doing that, we also estimate the offset in real-time in the coordinated direction. We have discussed all these steps in detail in the following subsections.

## 4.1. Vehicle platoon identification and count prediction

The first step of the CV-based adaptive signal control algorithm is to identify the vehicle platoons and estimate the number of CVs and non-CVs in the estimated platoon, which is discussed in the following subsections.

### 4.1.1 Vehicle platoon identification

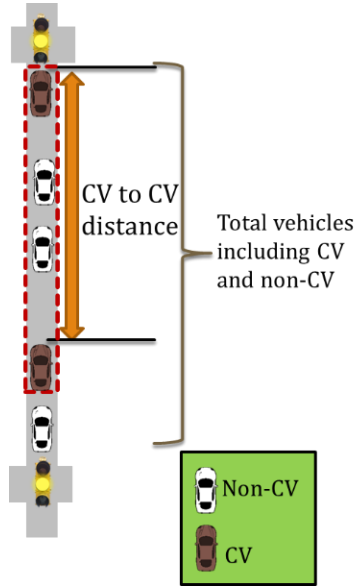

**Fig. 1** Platoon identification with CV and non-CV

The input of the adaptive signal system is the CV data and upstream traffic signal phase. Using only CV data and upstream signal phase, we update the traffic signal timing parameters (i.e., signal timing interval, sequence, and offset). At first, we identify the vehicle platoons based on spatial distribution of the CVs, as shown in Fig. 1 with the dotted bounding box. We define 'corridor segment' as the roadway section in each direction between two successive traffic signalized intersections (for major road), or the side street sections considered for this analysis (for minor road). On each single corridor segment, we identify platoons in two separate areas. The first one is the queued platoon which forms inside the expected queue zone, and the next one is the queued platoon which forms beyond the expected queue zone. The expected queue zone is the area near a traffic signal where queue is typically formed due to the signal yellow and red time. We locate this area either based on CV data when any CV is queued due to the traffic signal, or based on previous queue length data (i.e., based on historical observations) when there is no queued CV. If two vehicles are moving beyond the expected queue zone, we form a platoon using two successive CVs. Any non-CVs inside the two successive CVs will be considered as a part of the platoon. If there is only one CV, we only consider one platoon, which has only that CV. We similarly form platoons with CVs inside the expected queue zone, but we do not merge the platoons from both zones together.

### 4.1.2 Total traffic count prediction model

In order to estimate traffic signal timing for the next time horizon, we need to predict total traffic count (including both CVs and non-CVs) for a certain time horizon. Based on this total traffic count, we will optimize the intersection specific green time intervals and corridor-specific offsets. We predict the total traffic count for each corridor segment using the Long Short-term Memory or LSTM models. We have derived the LSTM batch size (i.e., the size of the data that will be forwarded together in the learning step), epoch number (i.e., defining how many times the entire dataset will go forward and backward through the network), neuron number, and weights initially from a large group of parameters for 20 corridors segments (randomly selected). Based on the results from 20 models, we have filtered the parameter group and made a small group to figure out the batch size and neuron number for each corridor-specific LSTM model for different CV penetration levels. Using the Python Keras (Chollet 2015) library, we have implemented the framework, and we have estimated the parameters with cross-validation using the Clemson University Palmetto Cluster computing infrastructure. Using the cluster-computing environment the computation time is significantly reduced, and the optimized parameters are derived. While deriving the optimal parameter values, we have used 70% of the training data to train, and 30% of the training data to validate the LSTM models. We have estimated the accuracy of different prediction time windows to figure out which time-window based prediction gives us the most accurate result.

We have used an iterative process which uses Eq. 1 to estimate the required number of data in the training set, assuming that the underlying data is normally distributed. Here n is the number of samples. In this process, we have estimated the sample size which satisfies the required number of sample runs, as shown in Fig. 2. Using Root Mean Square Error or RMSE, we have calculated the accuracy of the LSTM prediction as shown in Eq. 2. The low RMSE value means the predicted total traffic count (including both CVs and non-CVs) are very close to the actual total traffic count.

$$n = \left(\frac{z_{\frac{\alpha}{2}} * \sigma}{E}\right)^2 \quad (1)$$

$$RMSE = \sqrt{\frac{\sum_{i=1}^{T}(y_{i,act} - y_{i,pre})^2}{T}} \quad (2)$$

where, $Z_{\frac{\alpha}{2}} = 1.96$ (at a 95% confidence interval)
$\sigma$ = Standard deviation of simulation travel time
$E$ = Tolerance (i.e., 6% of the average travel time captured from the field)
$n$ = Number of simulation runs
$T$ = Total number of samples
$y_{i,act}$ = Actual data for $i^{th}$ observation
$y_{i,pre}$ = Predicted data for $i^{th}$ observation

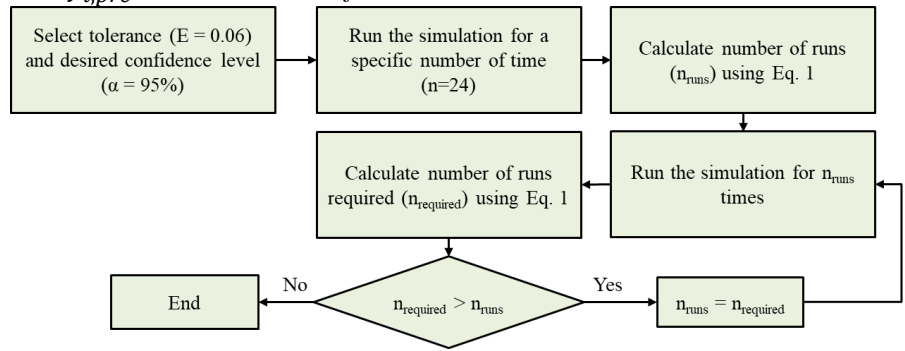

Fig. 2. Estimation of required sample runs.

*4.2. Real-time intersection signal timing parameters optimization*

In this section, we discuss the signal timing parameter estimation method for each intersection. First, we discuss the green interval estimation process, followed by a discussion on queue dissipation time based on shockwave theory. The queue dissipation time is used to identify which platoon will be affected by the existing queue in the intersection. Later, we discuss the minor street green time calculation step. The last subsection discusses the multi-objective optimization formulation step to estimate the intersection-specific signal timing parameters.

### 4.2.1 Traffic signal timing estimation

Based on the LSTM-derived total traffic count for each street, we have calculated the critical lane volume to derive maximum green time for non-coordinated phases. The following shows the equations, notations and values to estimate the signal timing parameters for both coordinated (e.g., phase 2 and 6) and non-coordinated phases (e.g., phase 4 and 8). For the CV-based adaptive signal control, we have only used coordination for one direction in the major street, which is the west-to-east direction (referred to as the coordinated direction in later sections) for the case study area. We have not adjusted offsets based on the traffic in the opposite direction (east-to-west direction, referred as the non-coordinated direction in later sections) on the major road. The maximum and minimum green time can be estimated based on different traffic engineering manuals and experience based on engineering practices (FHWA 2017; HCM 2010; MnDOT 2013; UDOT 2017). In this study, for the maximum green time in the non-coordinated phases, we have considered the minimum of the SYNCHRO calculated maximum green time for that phase and the required green time based on the real-time traffic condition in the side street. Based on the peak hour volume, the maximum green time can be calculated in SYNCHRO, which is a fixed value.

Coordinated Phases

$$g_{coord,\,max} = C - PC_{coord} - \sum_{i=1}^{P}(g_{non-coord,min,i} + PC_{non-coord,min,i}) \qquad (3)$$

Non-coordinated Phases

$$g_{non\text{-}coord,\,max} = min\left(g_{non\text{-}coord,max,fixed},\ (C\text{-}L)\frac{VC_{non-coord}}{VC}\right) \qquad (4)$$

where, $C$ = Cycle time (sec.)
$i$ = Non-coordinated phase
$P$ = Total number of non-coordinated phases
$PC_{non\text{-}coord}$ = Phase clearance time the non-coordinated phase (sec.)
$L$ = Total lost time (sec.)
$VC_{non-coord}$ = Critical lane volume for the non-coordinated phase (veh/hr)
$VC$ = Sum of critical lane volume (veh/hr)
$g_{coord/non\text{-}coord,\,max}$ = Maximum green time for the coordinated/non-coordinated phase (sec.)
$g_{non\text{-}coord,\,min}$ = Minimum green time for the non-coordinated phase (sec.) (we have considered 4 sec. as the minimum green for all phases)
$g_{non\text{-}coord,\,max,\,fixed}$ = Maximum fixed green time for the non-coord phase derived from manuals or traffic engineering software (sec.)

### 4.2.2 Queue dissipation time estimation for major approaches

For each signalized intersection, we need to estimate the queue dissipation time for both approaches in the major direction. After the green interval ends, vehicles in the major direction start to form a queue. If $L_Q$ is the maximum queue length (as shown in Fig. 3 (a)), with the following two logics we can estimate $L_Q$.

- If CV-based platoons exist close to the traffic signal, and the platoon speed is less than 5 mph, we consider the rear-point location of the CV-based platoons as the extent of $L_Q$.
- If no CV exists in the queued area in the major approaches due to the low-penetration level of CVs, we consider $L_Q$ for a certain approach as the typical queue length for that approach estimated based on the historical observations.

Once we know $L_Q$, we divide the corridor segment into two sections: congested section and uncongested section, as shown in Fig. 3 (b). We derive the jam density ($k_j$) for each approach based on the following Eq. 5.

$$k_a - k_j = \frac{Rq_a}{L_Q} \quad (5)$$

where, $R$ = Inter-green time (sec.)
$L_Q$ = Maximum queue length (mile)
$k_a$ = Uncongested section density (veh/mile)
$q_a$ = Uncongested section flow rate (veh/hour)

After estimating the jam density, we calculate the backward forming shockwave speed $(v_{bf})$ and backward recovery shockwave speed $(v_{br})$ using Eq. 6 and 7.

$$v_{bf} = \frac{0 - q_a}{k_j - k_a} \quad (6)$$

$$v_{br} = \frac{L_Q}{T_{Sq}} \quad (7)$$

where, $T_{Sq}$ = Time when last queued vehicle starts to move after the signal turns green (measured from the simulation case study: 1.5 seconds for each successive vehicles)

Once we get $v_{bf}$ and $v_{br}$, we calculate queue dissipation time $(T_Q)$ using the following Eq. 8. $T_Q$ cannot be more than the cycle time, as we do not expect to experience network spillover.

$$T_Q = \frac{R \, v_{bf}}{v_{bf} - v_{br}} \quad (8)$$

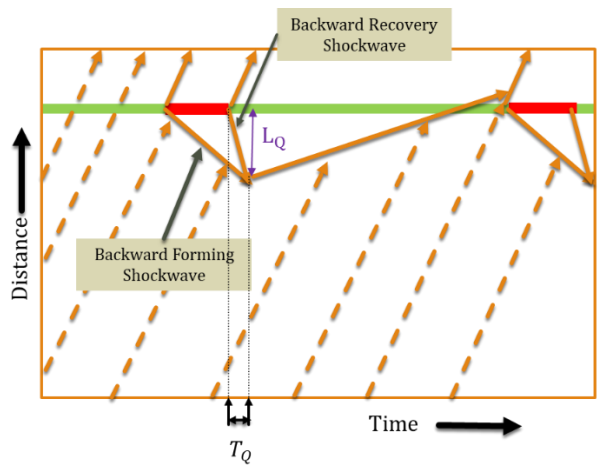

(a) Backward forming and recovery shockwaves for one intersection

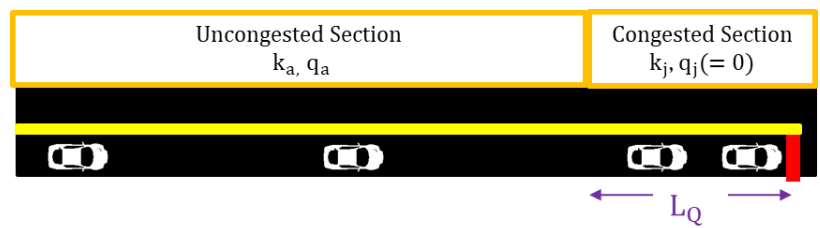

(b) Corridor sections based on $L_Q$

**Fig. 3.** Maximum queue length $(L_Q)$ and time to dissipate the queue $(T_Q)$ for one approach.

### 4.2.3 Minor road green activation time

Using Eq. 4, we have estimated the upper bound of the green time for non-coordinated phases. In order to initiate the green time for the minor approaches, we have to monitor the queue length on both sides of the minor approaches. As illustrated in Fig. 4, the minor direction queue starts to form when the signal is red in that direction. Based on historical observations, we have the maximum allowable queue length data for both sides of the minor approach. Here two scenarios could happen. The first scenario applies when CV-based platoons are present in the queue. By tracking the movement of the CVs, if moving platoons are found within the maximum allowable queue length, we do not initiate the green time for the minor approach. If we find queued platoons (based on CV platoons speed) beyond the length of the maximum allowable queue length, we initiate the minor direction green time. In the second scenario, if only non-CVs exist on the minor approach, we calculate queue length based on the predicted total traffic count on the side street for the amount of time passed from the start of the inter-green time. If we find that the number of vehicles queued for a certain time interval has exceeded the allowable queue length, we initiate the green time for the minor approach.

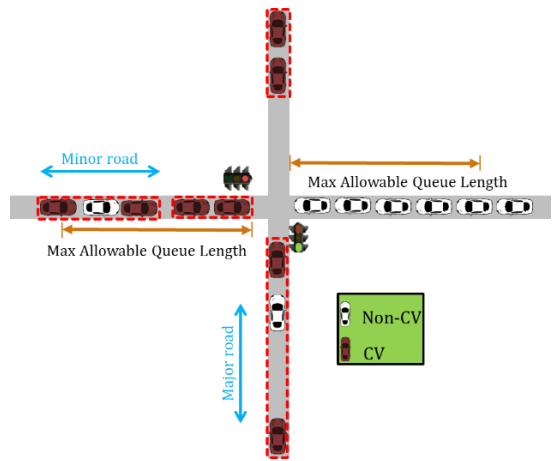

**Fig. 4.** Green activation time for minor approaches

### 4.2.4 *Multi-objective MILP formulation for intersection*

We have developed the multi-objective MILP formulation for each individual intersection to minimize the vehicle waiting time (due to the preexisting queue and inter-green time interval), and to maximize the number of progressing platoons. For this step, we are considering three different scenarios, which we discuss here.

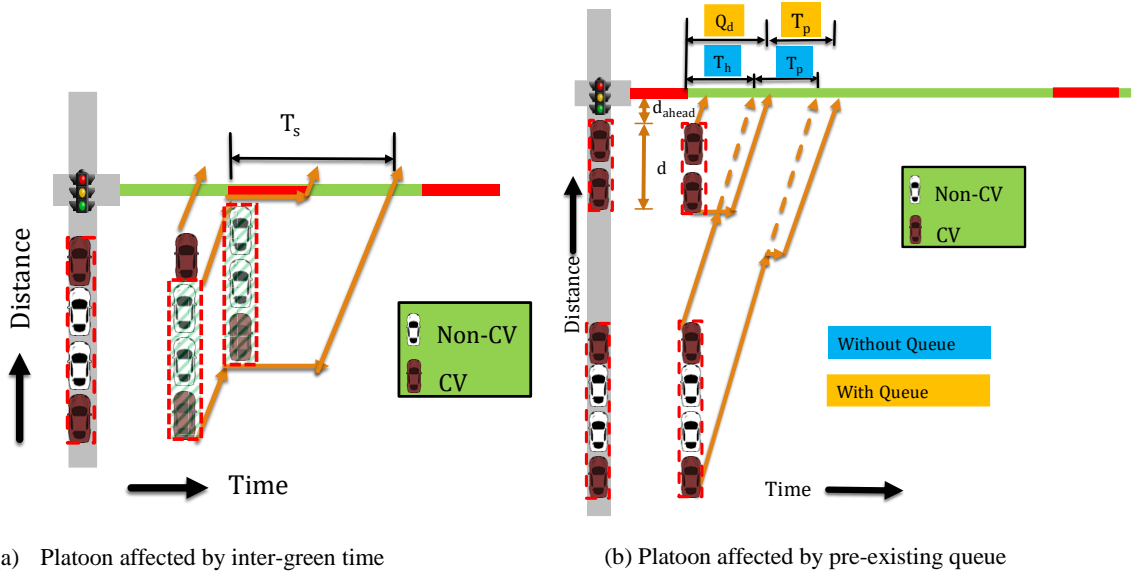

(a) Platoon affected by inter-green time

(b) Platoon affected by pre-existing queue

**Fig. 5.** Platoon progression affected by signal timing and pre-existing queue

In the first scenario, we consider vehicle-waiting time due to the inter-green interval in the major approaches. As shown in Fig. 5 (a), the platoon progression is affected due to the end of green time, as a fraction of the platoon cannot cross the intersection. The hashed bounding box in Fig. 5 (a) shows the inter-green time-affected portion of the platoon, which needs to wait for the inter-green time, and the time required to clear the portion of the platoon is $T_s$. We calculate $T_s$ using the following Eq. 9. Here $f_r$ is one of the decision variables of the MILP framework.

$$T_s = IG_{cor} + \sum_{i=1}^{M} \frac{d_{p_i}(1-f_{r_i})Nh_s}{L} \tag{9}$$

where, $f_{r_i}$ = Fraction of the $i^{th}$ platoon that is allowed to go
$h_s$ = Saturation headway (sec)
$d_{p_i}$ = Length of the $i^{th}$ platoon (mile)
$IG_{cor}$ = Inter-green time of coordinated phase (sec)
$N$ = Number of vehicles in the specific approach of corridor segment
$M$ = Total number of platoons in the specific approach of corridor segment
$L$ = Length of the corridor (mile)

The second scenario includes a situation where a platoon progression in the major approaches is affected by the pre-existing queue at the intersection stop-line during the green time. As shown in Fig. 5 (b), the approaching platoon could cross the intersection in a shorter time period (shown with blue colored time duration) if there is no platoon queued in the intersection, compared to the scenario when queued platoon exists (shown with yellow colored time duration). The following Eq. 10 (without queue consideration) and 11 (with queue consideration) show the equation to calculate the time to clear the queue due to pre-existing queue/congestion ($T_q$) for this second scenario:

$$T_q = T_{h_1} + T_{l_1} + \sum_{i=2}^{M} ((T_{h_i} + T_{l_i}) - (T_{h_{i-1}} + T_{l_{1-1}})) \tag{10}$$

$$T_q = Q_d + \sum_{i \in A} \sum_{i=1}^{M}(d_{p_i} f_{r_i} + d_{ahead})/V_{a_i} + \sum_{i \in B} \sum_{i=1}^{M}(T_{h_i} + T_{l_i}) \tag{11}$$

where, $V_{a_i}$ = Average speed of the $i^{th}$ platoon (mph)
$A$= Set of approaching platoons affected by the queue
$B$= Set of approaching platoons not affected by the queue
$d_{ahead}$=Distance of the front vehicle of a queued platoon from the intersection stop line (mile)

$g$=Green time for the major approaches (sec.)
$Q_d$=Time to clear the pre-existing queue (sec.)
$T_{h_i}$=Time required for the first vehicle of the $i^{th}$ platoon to reach the stop line if no queue existed (sec.)
$T_{l_i}$=Time required for the last vehicle of the $i^{th}$ platoon to reach the stop line (sec.)
$=(d_{p_i} f_{r_i})/V_{a_i}$

The third scenario is for the platoons, which are not allowed to pass the intersection during the current cycle. These platoons need to wait for the whole cycle time, and we need to reduce the waiting times required by these platoons. We can estimate the waiting time for these platoons ($T_p$) using the following Eq. 12.

$$T_p = \sum_{i=1}^{M} C \delta N_{non} \tag{12}$$

where, $N_{non}$= Number of vehicles in non-selected platoons

$$= \sum_{i=1}^{P}\left(\#CV_i + \frac{(d_i-10)N}{L}\right)$$

$\#CV_i$= Number of CV in the $i^{th}$ platoon

$$\delta = \begin{cases} 1 \text{ if } fr = 0 \\ 0 \text{ if } fr \neq 0 \end{cases}$$

Based on these three scenarios, the MILP multi-objective optimization model, which is developed, is shown in Eq. 13 and 14.

Multi-objective functions:

$$\text{Minimize} \sum(T_q+T_s+T_p) \tag{13}$$

$$\text{Maximize} \sum_{i \in A,B}^{M}(\#CV_i + \frac{(d_i-10)Nf_{r_i}}{L}) \tag{14}$$

Subject to:

Inequality Constraints:

$T_q \leq g$

$g_{min} \leq g \leq g_{max}$

$0 \leq f_{r_i} \leq 1$

Equality Constraints:

$g_{cor(2,6)} + y_{cor(2,6)} + r_{cor(2,6)} + g_{non-cor(4,8)} + y_{non-cor(4,8)} + r_{non-cor(4,8)} = C$

Decision Variables:

$f_{r_i}$, $g_{cor(2,6)}$, $g_{non-cor(4,8)}$

$g_{cor(2,6)}$, $g_{non-cor(4,8)}$ are integers

where, $g_{cor}$ = Green time for the coordinated approach
$y_{cor}$ = Yellow time for the coordinated approach
$r_{cor}$=Red time for the coordinated approach
$g_{non-cor}$ = Green time for the non-coordinated approach
$y_{non-cor}$ = Yellow time for the non-coordinated approach
$r_{non-cor}$=Red time for the non-coordinated approach

The optimization problem is solved in real-time based on the demand in the coordinated approach. As shown in Fig. 6, while the signal in the coordinated approach turns green from red, the optimization algorithm computes the green time required by the platoons (including CVs and non-CVs based on the predicted total vehicle count) to cross the corridor in the coordinated approach. It simultaneously computes the maximum green time required by the side street following Eq. 4. We use the concept of 'grace-period' for the major approaches, which means that we give a certain amount of green time to the major direction at the start of the green time. In case of limited CV penetration and/or limited traffic in the major direction for one single intersection, if we do not find a CV-based platoon to compute green (at the start of the green) in the side street, it can have an adverse impact on the coordination. Allowing green time for a certain grace period gives us the opportunity to wait for a certain time interval to find a CV-based platoon in the major approach, and perform the optimization. This grace period is equivalent to 'yield point' of actuated-coordinated signal, where the green is provided for a certain time period on the coordinated direction. If the green time required by the platoons in the coordinated direction is higher than the signal timing split with the maximum green time for the minor direction, we need to initiate the process of allowing side-street green time based on the side-street queue length (as discussed in Section 4.2.3). This simply means that we have provided the sufficient green time in the coordinated major approaches for a certain cycle length C, and now we need to monitor the side street queue to initiate the side street green time. However, if the green time required by the platoons in the coordinated direction is lower than the signal timing split with maximum green time for the minor direction or if there is no call from the side street, we can allow more green time to the major approaches. In this case, we look for the CV-based platoons in the major coordinated approaches for a certain time-period before the previously calculated green time ends (10 seconds considered in this research). If we get any platoon in that interval, we perform the optimization again to re-compute green time for both directions for the remaining cycle time. If we do not get any platoon, we initiate the side-street green. In any case, the green time not used by the non-coordinated phases is used by the phases in the coordinated direction. We have used MIDACO solver (Schlueter et al. 2013) to perform all MILP optimization. Using the multi-processing capability of MIDACO, we have reduced the real-time computation time.

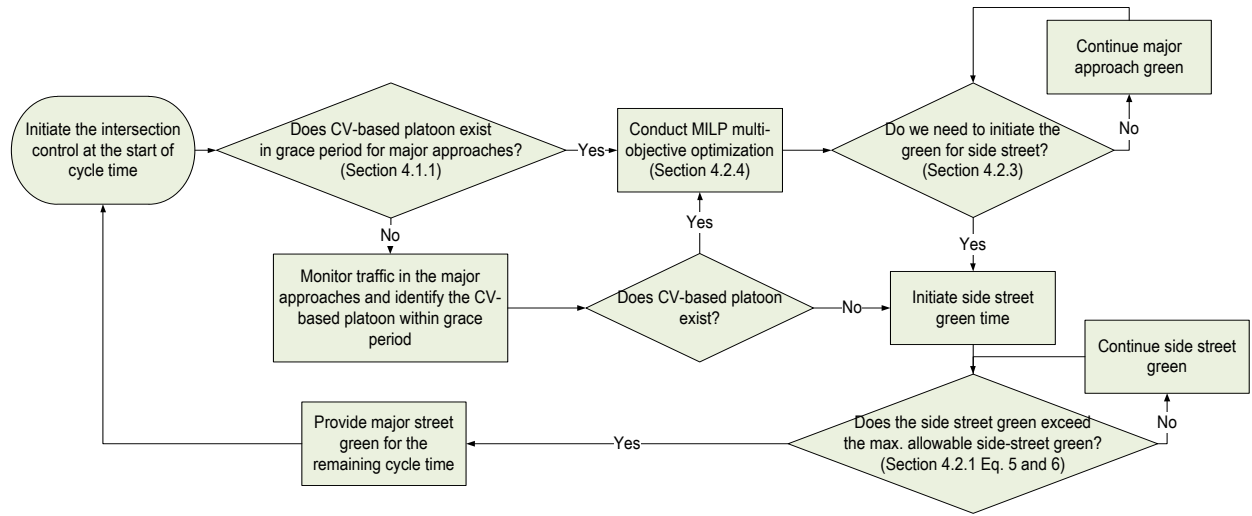

**Fig. 6.** Intersection signal control strategy

*4.3. Real-time offset optimization*

In this step, we discuss the offset optimization strategy for a corridor using Fig. 7. The purpose of this step is to distribute signal timing parameters in such a way that the platoons, which will start moving from an upstream intersection, will not face the queue in the downstream intersection. In order to compute the real-time offsets, we use the upstream vehicle flow rates from the side streets (i.e., $Q_{s1}$ and $Q_{s2}$ from two side streets), and the main street (i.e., $Q_{up}$ for upstream intersection and $Q_{dis}$ for downstream intersection), as shown in Fig. 7 (a).

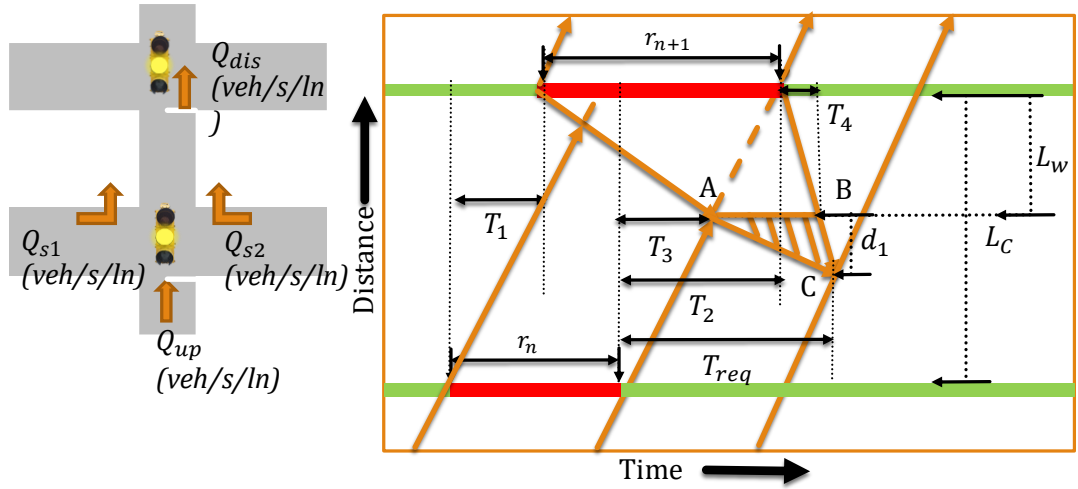

(a) Saturation flow rate        (b) Offset optimization

**Fig. 7.** Offset optimization for urban arterials.

As shown in Fig. 7 (b), the platoon, which will start from the upstream intersection, will face delay if these is a pre-existing queue in the downstream intersection, and that delay is due to the area of ABC. The MINLP formulation is shown in Eq. 15.

$$Minimize\ Delay = \frac{Area\ of\ ABC}{h} = \frac{AB*d_1}{2h} = \frac{(T_2+T_4-T_3)*(T_{req}Q_{up}h)}{2h} \quad (15)$$

Subject to:
    Inequality Constraints:
$$-\frac{L}{V_{FFS}*\alpha} \leq T_2 \leq \frac{L}{V_{FFS}*\alpha}$$
$$1 \leq T_1 \leq \frac{C}{2}$$
    Equality Constraints:
$$r_{n+1} = |r_n - T_1| + T_2$$
    Integer Constraints:
        $g_{cor},\ g_{non-cor}$
    Decision Variables:
        $T_1,\ T_2$ (integers)

where, $r_n$ = Inter-green time for the upstream intersection (sec.)
$r_{n+1}$ = Inter-green time for the downstream intersection (sec.)
$L_w$ = Distance between downstream intersection stop line and platoon arrival point to join the queue (mile)
$T_1$ = Time interval between the start of inter-green times for two intersections (sec.)
$T_2$ = Time interval between the start of green times for two intersections (sec.)
$T_3$ = Time interval between the start of upstream intersection inter-green time and platoon arrival time to reach the downstream queue (sec.)
$T_4$ = Time interval between the start of upstream intersection green time and platoon arrival time to reach the downstream queue (sec.)
$h$ = Measured saturation headway= 2.5 sec (from simulation $4^{th}$-$5^{th}$, $5^{th}$-$6^{th}$ vehicle)
$\alpha$ = User-defined factor (0.8 considered in our research)

Once the optimization is conducted, we derive the value of $L_w$ using Eq. 16. After estimating $L_w$, using the following Eq. 17 and 18, we calculate $T_3$ and $T_4$ respectively.

$$L_w = v_{bf} * ((|r_n - T_1|) + T_3) \tag{16}$$

$$T_3 = \frac{L - L_w}{V_{FFS}} = \frac{L - v_{bf}(|r_n - T_1|)}{V_{FFS} - v_{bf}} \tag{17}$$

$$T_4 = \frac{(|r_n - T_1|)(Q_{s1} + Q_{s2})}{Q_{dis}} \tag{18}$$

where, $Q_{dis}$=Downstream Discharge Saturation flow rate in veh/hour green/ln unit = $\frac{3600}{h}$veh/hg/ln

As we are not considering the impact of driveway traffic in order to simplify our analytical framework, the downstream intersection discharge rate will be affected only by the upstream intersection incoming flow rate (from both side streets as $Q_{s1}$ and $Q_{s2}$, and main street as $Q_{up}$). This situation is valid if there is not significant driveway traffic demand for a corridor. According to the flow conservation, we can derive the following Eq. 19. We consider $T_{req}$ as the time required to achieve the free flow condition in the downstream intersection, meaning we will not find any signal related queue in the major approach after this time. We calculate $T_{req}$ using Eq. 20.

$$(|r_n - T_1|)(Q_{s1} + Q_{s2}) + T_{req}Q_{up} = (T_{req} - T_2)Q_{dis} \tag{19}$$

$$T_{req} = \frac{(|r_n - T_1|)(Q_{s1} + Q_{s2}) + T_2 Q_{dis}}{Q_{dis} - Q_{up}} \tag{20}$$

## 5. Experimental Design

In this section, we have discussed the study corridor for our CV-based signal control evaluation. Also, we have included discussion for both actuated coordinated and adaptive signal control scenarios.

### 5.1. Study Corridor

In order to evaluate the performance of the CV-based adaptive traffic control framework, we have used the US 29 corridor from Greenville, South Carolina. This selected corridor has 10 intersections, and length of the study area is 3.2 miles. The yellow pin points in Fig. 8 shows the location of the signalized intersections along the US 29 corridor. We have simulated the corridor using the 'Simulation of Urban Mobility' (SUMO).

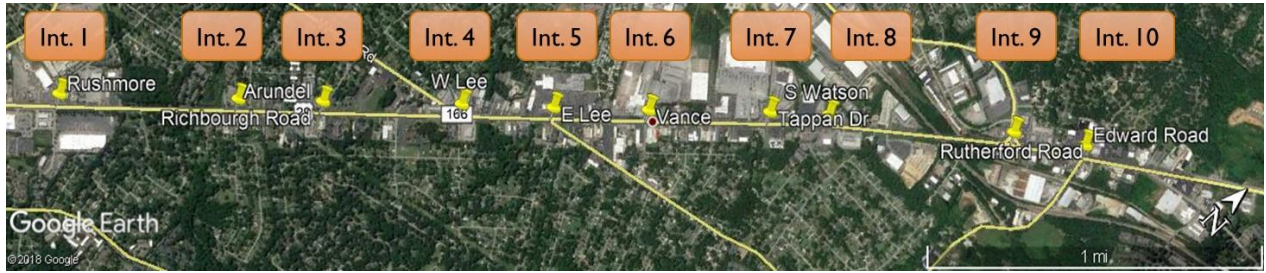

**Fig. 8.** US 29 Wade Hampton study corridor from Greenville, SC

We have calibrated the simulation network with turning traffic counts from each intersection, and travel time from peak hour traffic. For six intersections, we have collected historic turning traffic counts for the afternoon peak hour from South Carolina Department of Transportation (SCDOT). The intersections include Rushmore Drive, Richbourg Drive, Arundel Road, East Lee Road, South Watson Road, and Edward Road. We have collected turning counts for the other intersections and travel time for both directions on the major corridor. Apart from collecting data for the four other intersections (i.e., West Lee road, Vance Street, Tappan Road and Rutherford Road), we have collected data from East Lee and Edward Road to estimate the annual growth factor for the turning traffic, comparing the SCDOT provided data and our field collected data. We have estimated the annual growth factor as 1.25%, and applied this factor to estimate the turning traffic counts for the six intersections where we have used data from SCDOT. We have collected the existing signal timing plans from SCDOT to calibrate the model. We have calibrated the simulation model so that the travel time from the simulation models resides within 10% of that from the real-world data. We have used the Intelligent Driver Model (IDM) as the car-following model for the vehicles in SUMO. For the urban corridor, we have these parameters for the IDM model: Minimum Gap = 2m, Acceleration = 1 m/s$^3$, Deceleration = 1.7 m/s$^3$, Time Headway = 0.5 s, and Acceleration Exponent = 4.

Once the simulation model is calibrated, we have increased the peak hour traffic for all intersections to achieve a congested scenario. We have increased the intersection turning volumes so that the approach Level of Service for each intersection becomes C or worse, and the volume-by-capacity ratio becomes higher than 0.9. We have not achieved such an operational condition for Vance Street and the North Watson eastbound approach on the major street direction, and the Rutherford road major street direction, as the side street performance already deteriorates. Also, we have used only a two-phase based signal control system for simplification, as discussed in Section 4.2.1. However, the model can be applied to any scenario with more than two-phase signal control.

### 5.2. Actuated coordinated signal control

In order to evaluate the CV-based signal control's performance, we have considered the actuated-coordinated signal control time from SYNCHRO software as the baseline. SYNCHRO is widely used, both by public and private agencies, to estimate traffic signal timing parameters. For the baseline scenario with SYNCHRO-based actuated-coordinated control, the coordination is done for both directions on the major road (for both phase 2 and 6). The cycle time, phase splits, maximum green time, and offset values for each intersection are optimized using SYNCHRO. In this scenario, all intersections have the same cycle length. Using the inductive loop detectors, the minor direction green is provided only if vehicles are detected in the minor direction. Otherwise, the green signal resumes in the major direction. We have used fixed signal offset values in this scenario, based on the offsets derived from SYNCHRO.

### 5.3. CV-based adaptive signal control evaluation

To evaluate the CV-based adaptive signal control's performance, we have considered two-phase signal control for an urban arterial without any protected or exclusive left-turn phase. However, in the future we can add additional phases in the signal control framework to evaluate the signal control's performance in more complex scenarios. As shown in Fig. 9, Phase 2 and 6 are the phases for the major approach, and Phase 4 and 8 are the phases for the minor approach or side streets. The traffic signals are coordinated in the major approach. We have considered two-phase signal control only for simplification, and the algorithm can accommodate any higher number of phases for any given signal controller.

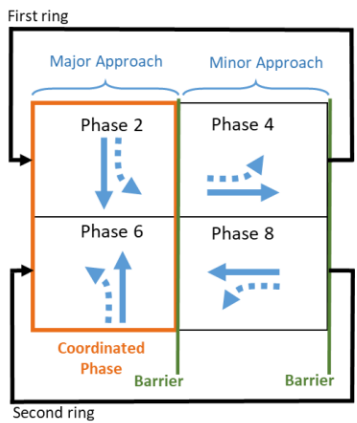

**Fig. 9.** Ring-barrier diagram for the two-phase intersection control.

In order to compare the performance of different signal controls, we have used the average of speed, maximum queue length, and stopped delay as the measures of the effectiveness. Using the lane area detectors in SUMO, we have measured the performance of different signal controls. In reality, the lane area detector is equivalent to a vehicle-tracking camera, where we track the vehicles for a certain length along the corridor, specified by start and end points. These detectors are not used by any traffic signal control strategy. The actuated coordinated signal control uses inductive loop detectors separately placed close to the intersections, while the CV-based signal control only uses CV-based data, and upstream signal phase. Using these lane area detectors, SUMO captures the length of the longest queued section during each time step to provide the average of the maximum queue length. With this measure, we can compare the change in average queue length in the adaptive signal control scenario compared to the actuated signal control scenario.

6. Analysis and Discussion

The following subsections discuss the study findings. We have used a short-term traffic forecasting model, and we have discussed the findings regarding the model performance in the first subsection. The following subsections discuss the evaluation of the CV-based adaptive signal control.

*6.1. Total vehicle count prediction*

Based on Eq. 1, we have found that the required number of samples for our testing is 32. For the LSTM-based total traffic count prediction, we have considered 32 training files and 8 test files, each with 1-hour of data. We have generated these all files using the SUMO software with different random seed numbers for different penetration levels of CV. We have discarded the initial 900 seconds of the simulation as the simulation warm-up time. We have derived the LSTM hyperparameters (i.e., number of epoch and batch size), layer number and neuron number in each layer based on the cross-validation accuracy or RMSE value using a 70-30 split of the training data (i.e., data from 32 training files) as the training-validation data set. For the prediction of the next time interval, the LSTM input parameters (from each corridor segment) for a certain second are: CV to CV distance (as shown in Fig. 1), CV speed, CV Waiting time, Number of CVs, Upstream Signal Phase (if an upstream intersection signal exists, otherwise not considered) and CV Direction. We considered the upstream signal phase based on majority voting from a previous time window. If we consider a 5-second prediction time horizon, the majority voting system will help us to identify the input signal phase that occurred most of the time in that 5 second interval. Based on Pearson's correlation coefficient, we have found that the input parameters are not correlated. The LSTM model output is the total traffic count (including both CVs and non-CVs) for the next time interval. After multiple iterations, it is found that the Nadam optimizer works best for the total traffic count forecasting, and the parameters used in this research for the Nadam optimizer are: learning rate=0.001, epsilon=None, schedule decay=0.004, and exponential decay rates, beta_1 and beta_2, are 0.9 and 0.999, respectively.

We have estimated the accuracy of the prediction time window, and found that the 1 second prediction accuracy is higher than the accuracy with the 2s, 5s, or 10s prediction windows. This small prediction time window allow us to check the traffic condition at a very fine level, as in reality traffic is very dynamic, and the situation can change highly if the prediction time window is high. Fig. 10 shows the performance of the LSTM-based prediction models for the major approach. Here using the optimized LSTM parameters, we show the RMSE from 8 test files for all corridor segments on US 29. With the increasing number of CVs, the RMSE value decreases. The RMSE with 5% CV penetration is 9.76, and with 100% CV is 5.66.

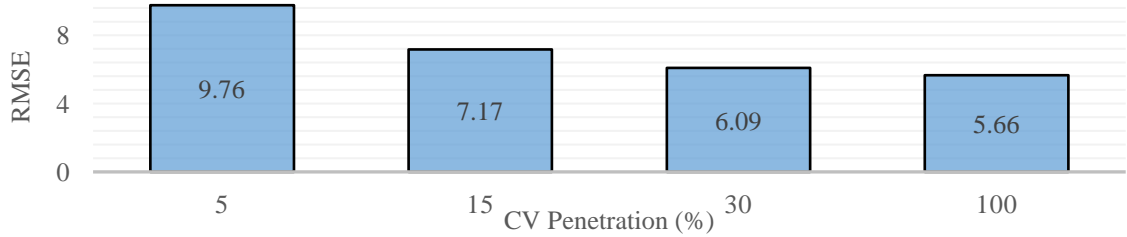

**Fig. 10.** LSTM performance of total count prediction for the next second.

Fig. 11 shows the LSTM model's performance for both 30% and 100% CV penetration for one corridor segment. At 30% CV penetration, the predicted total vehicle count (dotted line in Fig. 11(a)) is sometimes higher and sometimes lower than the actual count (solid line in Fig. 11 (a)). For 100% CV penetration, the predicted total count (dotted line in in Fig. 11(b)) is closely following the actual count (solid line in in Fig. 11(b)).

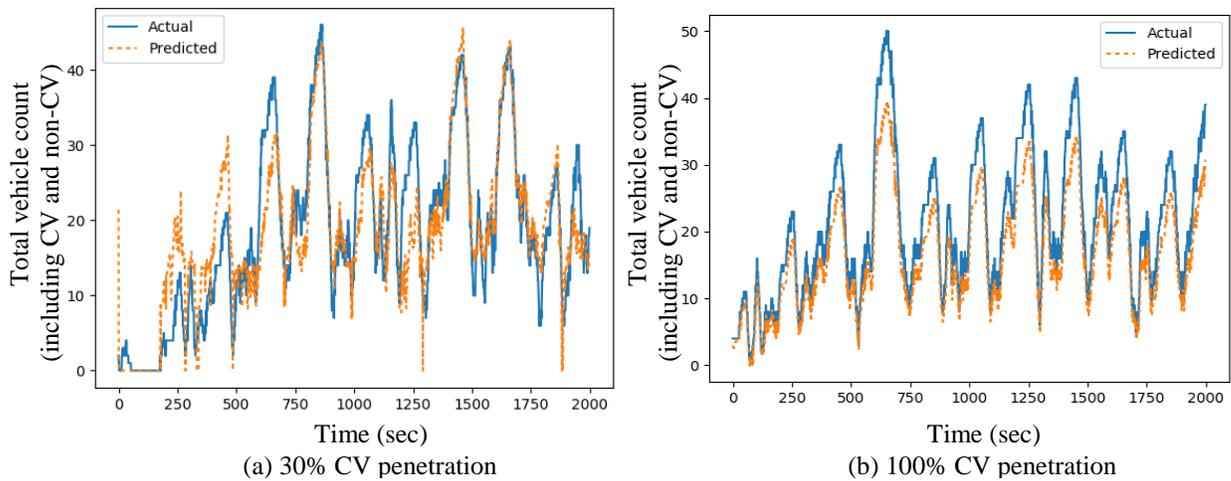

(a) 30% CV penetration  (b) 100% CV penetration

**Fig. 11** Total count prediction for different CV penetration

### 6.2. Evaluation of CV-based adaptive signal controls

We have included discussion about the impacts of CV-based adaptive signal control on both major and minor streets.

#### 6.2.1 Major Street Impact Evaluation

We have conducted the operational performance evaluation for all 10 intersections (as shown in Fig. 8) using both actuated coordinated signal control and adaptive signal control. In the adaptive signal scenario, the green intervals for both major and minor directions are dynamically decided in real-time based on intersection approaching vehicles. The offset timings are also adjusted for each cycle so that the approaching vehicles from one intersection do not need to stop at the following intersection. We will demonstrate the benefit of this system with the time-space diagram in Fig. 12. Fig. 12 shows the trajectory of the through traffic platoon throughout the corridor for both the actuated-coordinated and adaptive signal control scenarios. For the adaptive coordinated direction (orange arrow trajectory in Fig. 12) with 5% CV, we find the intermediate signal timings are synchronized in such a way that the platoon can reach the last intersection with a minimum interruption (#1 dotted box in Fig. 12). The #2 dotted box in Fig. 12 shows that with the fixed offset value in actuated coordinated control, the end section of the platoon reaches the last intersection after a longer time compared to the adaptive scenario. As the offset in the coordinated direction does not adjust in real-time, the platoon breaks down almost at the middle of the corridor (at intersection 4, shown with #5 box in Fig. 12). In the opposite direction (meaning approaches with phase 6 which are not considered for adjusting offset), the end section of the platoons reach the end intersection almost at the same time (comparing the #3 and #4 boxes in Fig. 12).

Fig. 13 shows the summary of the operational analysis for the actuated coordinated and CV-based adaptive signal. We have findings for both adaptive coordinated direction (i.e., approaches with phase 2) and opposite direction (i.e., approaches with phase 6), and the findings include the result for both CV and non-CVs. The numbers on the top of the bar charts show the average of 32 runs, and the bold numbers in percentages inside the bars show the relative increase (shown as '+') and decrease (shown as '-') with different CV penetration levels compared to the actuated coordinated scenario. The analysis shows that with increasing number of CVs in the traffic stream, the average speed increases, and average maximum queue length and average stopped delay are reduced in the coordinated direction. With only 5% CV data, we derive 5.6% average speed increase, and 66.7% and 32.4% reduction in average maximum queue length and stopped delay, respectively, in the major direction compared to the actuated coordinated scenario. With 100% CV penetration, we get 8.1% average speed increase, and 70.2% and 41.4% reduction in average maximum queue length and stopped delay, respectively, in the major direction compared to the actuated coordinated scenario.

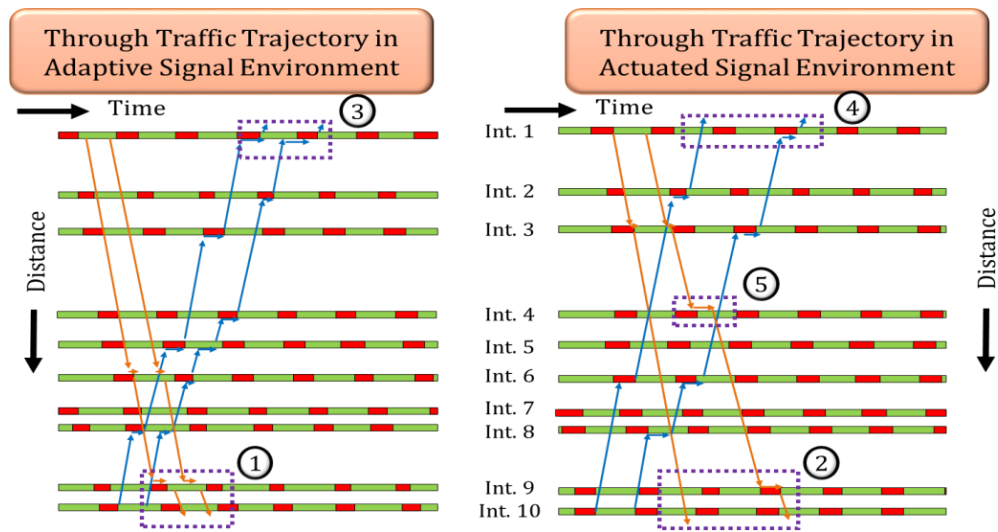

**Fig. 12** Trajectory of CV-based platoons for 5% CV penetration

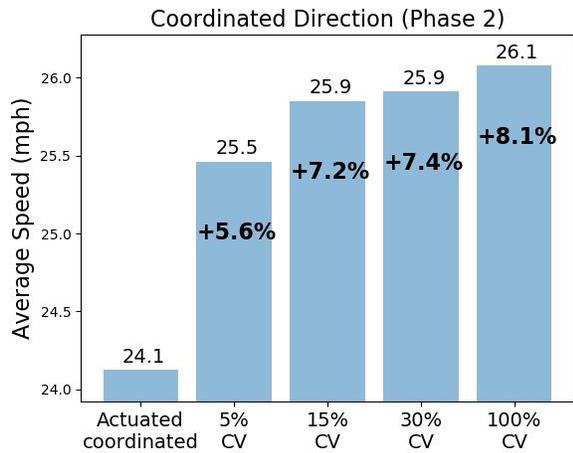
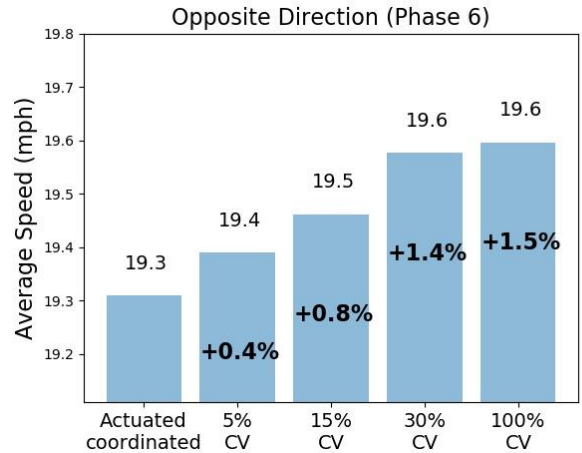

(a) Average speed comparison

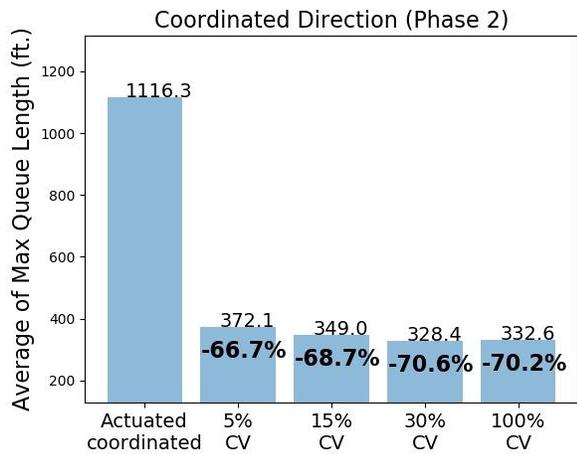
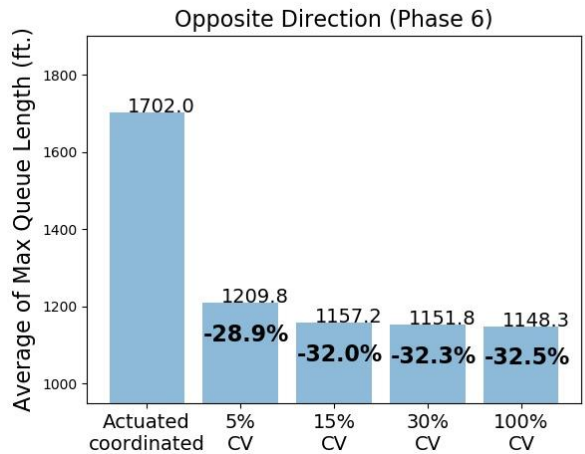

(b) Average of maximum queue length comparison

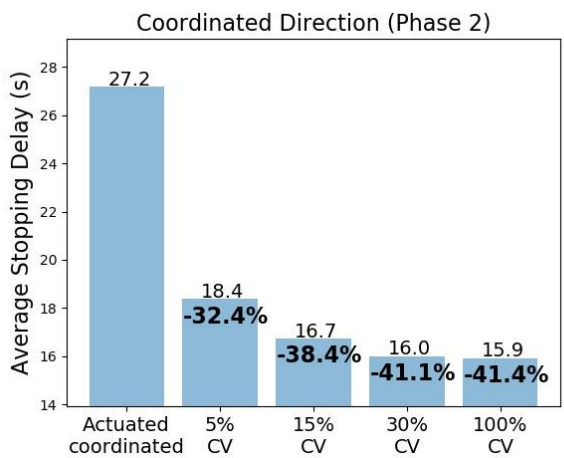
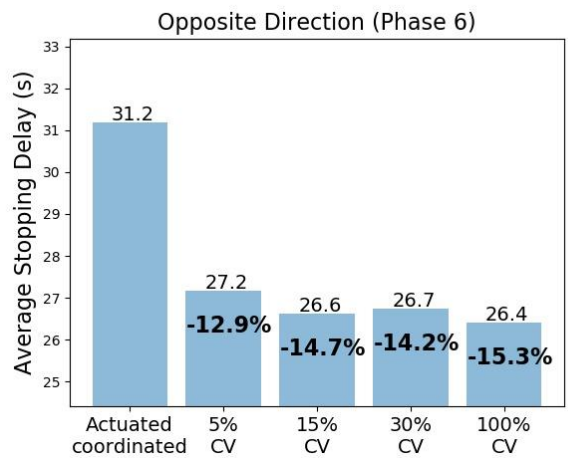

(c) Average stopping delay comparison

**Fig. 13** Summary operational impact findings for major streets

For the opposite direction (i.e., approaches with phase 6), the improvement is not as high like as those in the coordinated direction; however, we still find relative improvement in the traffic operational condition compared to the actuated coordinated scenario. With only 5% CV data, we achieve 0.4% average speed increase and 28.9% and 12.9% reduction in average maximum queue length and stopped delay, respectively, in the major direction with phase 6 compared to the actuated coordinated scenario. With 100% CV penetration we find 1.5% average speed increase and 32.5% and 15.3% reduction in average maximum queue length and stopped delay, respectively, in the major direction with phase 6 compared to the actuated coordinated scenario.

In summary, the CV-based adaptive signal control improves the operational condition of the corridor for both directions (i.e., adaptive coordinated direction with phase 2 and the opposite direction with phase 6) in the major street. Also, using only CV data, the operational performance can be improved even for very low CV penetration (5% CV). We can even provide benefits to the non-CVs with the limited data from 5% CVs.

### 6.2.2 *Minor Street Impact Analysis*

We have evaluated the impact of CV-based adaptive signal control on minor street/side street direction using the average speed. We have found that only 2 minor street approaches have experienced average speed increase among the total 18 minor street approaches in 10 intersections with 5% CV, whereas with 100% CV, only 1 side street approach has experienced the speed improvement. Our analysis reveals that the CV-based adaptive signal control does not improve the operational condition for all minor streets. Also, with increasing penetration, there is no trend observed in the effect on minor street traffic speed improvement.

## 7. Conclusions

With increasing connectivity and emerging digital infrastructure (i.e., connected infrastructure with communication and computation capabilities), we can utilize CV data to detect CV-based platoons and reduce our reliance on legacy transportation sensors, such as inductive loop detectors or video cameras, to detect approaching vehicles. The contribution of this research is unique as we have utilized only CV data and upstream signal phase information with an adaptive signal control strategy to improve a traffic signal controller's performance. Also, by determining the green interval and dynamically adjusting the intersection offset in real-time, we have achieved better operational performance compared to the traditional loop-detector based actuated-coordinated traffic signal. The CV-based adaptive signal control allows green time extension on the coordinated direction to satisfy the additional green time requirement in the case of fluctuating traffic demands. With LSTM models, we can reliably predict the total number of vehicles on certain corridor segments (i.e., the roadway section in each direction between two successive traffic signalized intersections for the major road or the side street sections considered for this analysis) for any penetration levels of CV. Once we know the information about the approaching CV-based platoons including non-CVs, we can proceed to estimate the intersection signal timing parameters. Based on our analysis, with only 5% CV data, we derive 5.6% average speed increase and 8%, 66.7% and 32.4% reduction in average delay, maximum queue length and stopped delay, respectively, in the major coordinated direction compared to the actuated coordinated scenario. With 100% CV penetration we obtain 8.1% average speed increase and 13.4%, 70.2% and 41.4% reduction in average delay, maximum queue length and stopped delay, respectively, in the major direction compared to the actuated coordinated scenario. Operational improvements occur in the major street opposite direction as well. However, for the minor streets, we have not found any distinct pattern regarding the impact of the CV-based adaptive signal control system.

In the future, the CV-based adaptive signal control algorithm can be extended to incorporate more complex scenarios on urban arterials. We will test the algorithm's performance while including more traffic phases (e.g., protected left-turn phase, pedestrian phase). Also, the different objective functions (e.g., energy efficiency improvement, safety improvement) can be included to enhance the algorithm. This CV-based adaptive signal's performance is validated using a calibrated simulation network. In the future, field validation should be conducted to evaluate the performance of the adaptive signal system in the real-world.

8. Acknowledgement

The authors would like to thank Aniqa Chowdhury for editing the paper.